\documentclass[12pt]{article}
\usepackage{amssymb}
\usepackage[dvips]{graphicx}
\begin{document}

%
%
%
%
\begin{center}
{\bf Calculations of the A$_1$ phonon frequency in photoexcited Tellurium}

P. Tangney and S. Fahy

Department of Physics,
University College, Cork, IRELAND.

{\bf ABSTRACT}
\end{center}

Calculations of the A$_1$ phonon frequency in photoexcited tellurium are 
presented. The phonon frequency as a function of photoexcited carrier density 
and phonon amplitude is determined. Recent pump probe experiments are interpreted 
in the light of these calculatons.
It is proposed that, in conjunction with measurements of the phonon period 
in ultra-fast pump-probe reflectivity experiments, the calculated frequency 
shifts can be used to infer the evolution of the density of photoexcited 
carriers on a sub-picosecond time-scale.

%
%
%
\newpage
During the last decade, continuing advances in ultra-fast laser technology 
have made possible time-domain studies of coherent phonons on ever 
shorter time scales \cite{shah, merlin}. 
In pump-probe experiments it has been possible to excite a high 
density of electrons in a semiconductor on a timescale much smaller 
than the lattice vibrational period and detect the resulting motion 
of the ions by observing oscillations in the optical reflectivity \cite{hunsche}. 
In certain materials, 
femtosecond laser pulses may be used to excite electrons 
from bonding into antibonding states, thereby depleting the bond 
charge and changing the equilibrium positions of the atoms;
the atoms then oscillate about their new equilibrium --- a 
mechanism know as  ``displacive excitation of coherent phonons'' 
(DECP) \cite{zeiger}.  
At very high excitation densities, the bonds may be weakened and
the resulting motion has a frequency lower than that observed 
in conventional Raman scattering.

In recent femtosecond pump-probe experiments, 
Hunsche {\it et al.} \cite{hunsche} reported a linear 
decrease of the tellurium \(A_{1}\) phonon frequency and an 
increase of the phonon amplitude with pump fluence. 
Their results show a cosine time-dependence of the 
optical reflectivity oscillations as a function of pump-probe delay,
which is characteristic of a displaced equilibrium. 
Moreover, they observed that at high excitation densities the periods of 
successive single cycles of the oscillations after the 
pump-excitation tended to decrease.

In this Letter, we calculate from first principles the dependence 
of the Te \(A_{1}\) phonon frequency on the density of excited
electrons, thus opening the possibilty of using the measured
shift in phonon frequency to probe local electron-hole plasma
density on a sub-picosecond timescale. 
We calculate the dependence of frequency and equilibrium 
bondlength on excitation density. 
We also calculate the initial phonon amplitude as a function
of excited electron density, within the DECP mechanism.
Finally, we calculate the amplitude dependence of the phonon
frequencies and predict a significant anharmonicity of
the \(A_{1}\) phonon at very high carrier densities. 

If a laser pulse of duration much shorter than the timescale 
of the ionic motion is incident on a crystal, then an effectively 
instantaneous photoexcited carrier distribution is created.
Initially, the electron and hole distributions will be
non-thermal but carrier-carrier scattering \cite{lin1,lin2,lin3} 
will result in a thermal (Fermi-Dirac) distribution of electrons 
and of the holes in much less than 100 fs, the time-scale
of phonon motion. 
However, electron-hole recombination, resulting in a common 
chemical potential for electrons and holes, occurs on a time-scale
of approximately 1 ns, which is much longer than the phonon period.
Although some of the carrier energy will be lost to the lattice 
by carrier-optical-phonon scattering, we will assume that, within 
the timescales considered here (less than several ps), this 
effect is negligible. 

With these physical time-scales in mind, we assume in the present calculations
that follow that no electron-hole recombination occurs during phonon 
oscillations.
There will then be a constant density of electrons in the 
conduction bands (with an equal density of holes in the valence bands). 
A Fermi-Dirac distribution of each carrier type is assumed but
the chemical potential for electrons $\mu_{e^-}$ and for 
holes $\mu_{h^+}$ are not equal.
(Previous theoretical investigations 
\cite{biswas1,biswas2,stampfli,silvestrelli} 
of short-time phonon dynamics in systems subjected to electronic 
excitation by ultrashort laser pulses have assumed, for technical 
reasons, that the chemical potential for electrons $\mu_{e^-}$ and for 
holes $\mu_{h^+}$ are equal; this is physically equivalent to 
assuming that the electron-hole recombination time is much shorter 
than the phonon period.)

The stable form of tellurium at low pressure is $\alpha$-Te, 
in which two-fold coordinated tellurium atoms form
infinite helical chains parallel to the ${\bf c}$-axis of the
trigonal P3\(_{1}\)21-D\(_{3}^{4}\) structure \cite{donohue}.
The three atoms per unit cell are at 
(x0\(\frac{1}{3}\),0x\(\frac{2}{3}\),\(\overline{x}\overline{x}\)0)
and form a single, complete turn of a helical chain. 
Each helix is surrounded by six equidistant helices and each atom
has four second-nearest neighbours in these adjacent helices. 
The atomic position free parameter $x$ is equal to the ratio
of the radius of each helix to the interhelical distance.

The experimentally determined equilibrium value of \(x\) 
is \(x_{equil}\) = 0.2633 $\pm$ 0.0005  
and the room-temperature lattice
constants are a = 4.4561 \AA\ and c = 5.9271 \AA. 
At these values, the bond length is 2.834 \(\pm\) 0.002 \AA, 
the bond angle is 103.2 \(\pm\) 0.1\(^\circ\), 
and the second-nearest neighbour distance is 3.494 \AA\ \cite{donohue}.

The motion of atoms in the A\(_{1}\) phonon mode corresponds 
to a variation of the helical radius \(x\), maintaining the 
symmetry of the crystal.
We note that the equilibrium value of \(x\) is not determined
by symmetry, and so may be expected to change in the photoexcited
material, allowing DECP excitation of this mode.

At the special value, \( x = \frac{1}{3}\), 
the nearest and second-nearest neighbour distances become equal,
the atomic coordination number increases to six, the helical
chain structure is destroyed, and  
the structure can be classified as rhombohedral, with space group 
R$\overline{3}$m-D$^{5}_{3d}$ and one atom per unit cell.  
The high-pressure $\gamma$-Te form of tellurium, which is stable above 
\(\sim70\) kbar, has this structure \cite{donohue}.
The $\alpha$-Te structure may be viewed as a Peierls distortion
of the $\gamma$-Te structure, as we shall see below.

We have calculated structural total energies as a function of 
the \(A_{1}\) symmetry phonon displacement \(x\) and determined 
the vibrational frequency using the ``frozen phonon'' 
method \cite{pickett}. 
The total energy is calculated using standard density functional 
theory (DFT) methods, but with a constrained total occupation of 
the conduction bands, representing a semiconductor at a fixed 
electron-hole plasma density.
Although this represents a departure from the usual finite-temperature
DFT, it is a physically appropriate representation of the system
on time-scales much less than the electron-hole recombination time.
In general, the use of LDA band energies for excited states
is complicated by the fact that the true (quasi-particle)
excitation energies are not equal to the DFT eigenvalues.
However, in this system the LDA gap (0.26 eV) is not too much 
different from the experimental band gap (0.33 eV \cite{galli}) and the band 
dispersions within the LDA conduction and valence are very similar 
to the experimentally determined dispersions \cite{us}.
Thus, the approximation of using the LDA excited bands to calculate
the total energy of the photoexcited system is reasonably accurate
in the present context.

We have calculated the \(A_{1}\) phonon frequency for 
values of the electron-hole plasma density corresponding to a 
range \(0\%-1.25 \%\) of the valence electrons excited into the 
conduction bands, and for a range of electron temperatures
up to $k_BT = 0.2$eV.
We find the electron temperature has a negligible effect
on the phonon frequency.
This insensitivity to temperature suggests that the results we 
have obtained are not strongly dependent on the assumption
of a thermal distribution of carriers; the dominant effect on
the phonon frequency is simply the number of carriers excited
to the conduction (anti-bonding) states, and not their precise
distribution within those states.

Total energy calculations were performed 
using the plane-wave pseudopotential method of Ihm, 
Zunger and Cohen \cite{ihm}, with Hamann, Schl\"{u}ter 
and Chiang \cite{hamann} pseudopotentials and the 
local density approximation (LDA) 
exchange-correlation potential of Ceperley and Alder \cite{cep}. 
Plane waves up to energy cutoffs of 16 Ryd and 49 Ryd were
used in the expansion of the wavefunctions and the charge 
density, respectively. 
Self-consistent potentials, total energies and electronic 
densities of states were calculated using a regular 10x10x10 grid 
in the full Brillouin zone (124 k-points in the irreducible zone).
States in the valence bands (i.e., bands 1-9) at each k-point
were weighted by the Fermi-Dirac distribution with a temperature
$T$ and a chemical potential $\mu_{h^+}$. 
The value of $\mu_{h^+}$ is chosen to ensure the correct number 
of holes in the valence bands, corresponding to the appropriate
electron-hole density.
Similarly, the states in the conduction bands (i.e., bands 10 and
above) were filled according to the Fermi-Dirac distribution with
temperature $T$ and chemical potential $\mu_{e^-}$.
In all calculations, we have used the experimentally determined 
room temperature lattice constants.

\begin{figure}
\begin{center}
\includegraphics[width=13cm,height=13cm]{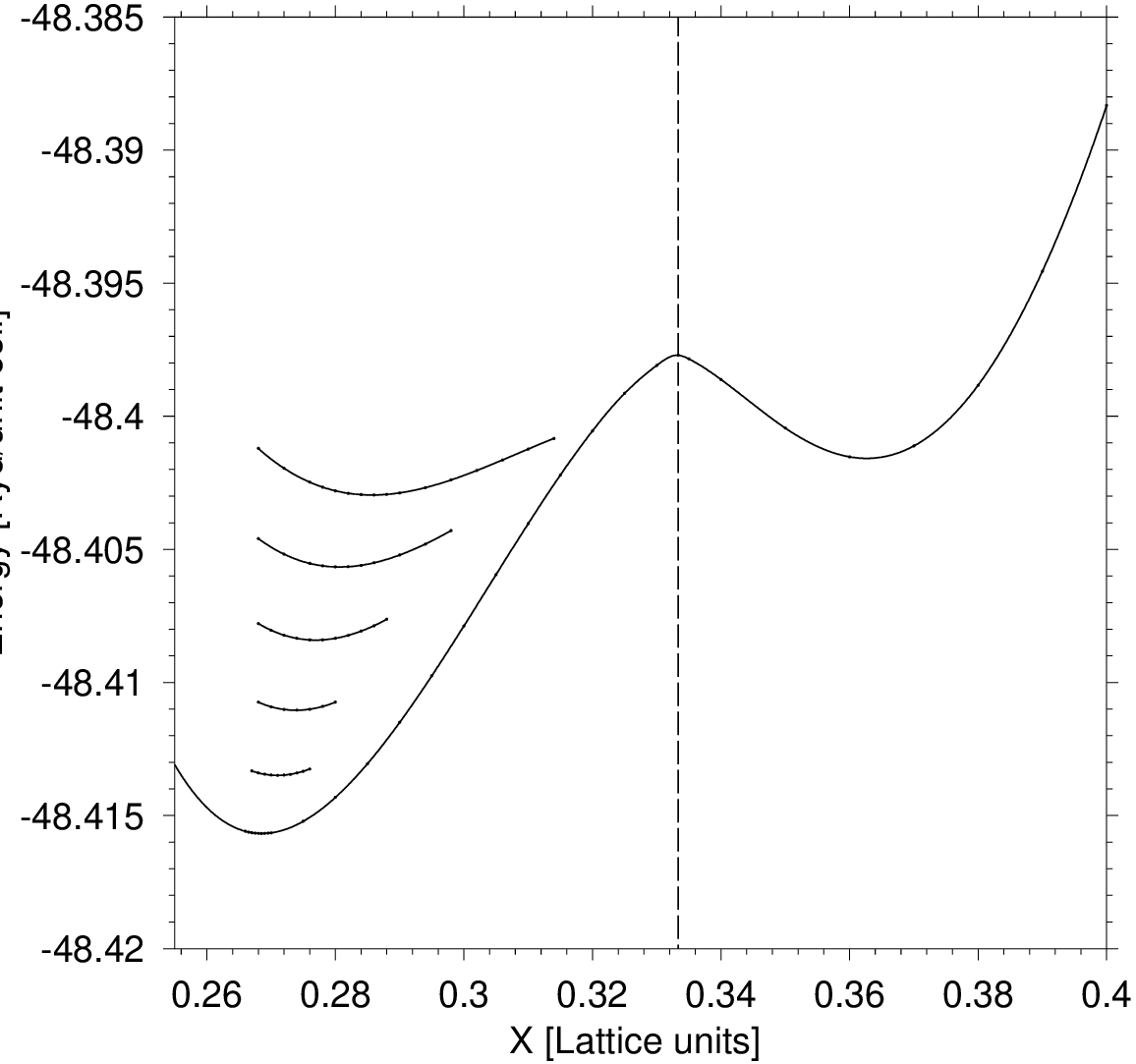}
\caption{Total energy per unit cell as a function of phonon coordinate %
\(x\) for photoexcited electron densities of \(0\%\) (lowermost curve), %
\(0.25\%\),\(0.5\%\),\(0.75\%\),\(1.0\%\) and \(1.25\%\) %
(uppermost curve) of the valence electron density.  %
The dashed line indicates the value of $x = \frac{1}{3}$ for the   %
high symmetry $\gamma$-Te structure.}
\end{center}
\end{figure}

Fig.\ 1 shows the structural energy as a function of \(x\) for 
excited carrier densities ranging from \(0\%\) to \(1.25\%\) 
of the valence electron density. 
The ground state equilibrium value of $x$ is found to be 0.2686, 
approximately \(2\%\) larger than the experimental value of 0.2633. 
The frequency of oscillation about this minimum was calculated 
as 3.24 THz, \(10\%\) smaller than the value obtained in 
conventional Raman scattering of 3.6 THz. 
This accuracy is typical of LDA calculations for such materials
\cite{louie,galli}.

The maximum at $x = \frac{1}{3}$ in Fig.\ 1 is due to a high
degeneracy of states near the Fermi level for this high
symmetry structure.
This degeneracy is lifted for $x \ne \frac{1}{3}$, lowering 
the energy of the occupied electronic states and the structural 
energy, in a striking example of a Peierls (or static Jahn-Teller) 
distortion. 
The high sensitivity of the A$_1$ phonon mode to photoexcitation
is a direct consequence of this mechanism for stabilization of
the $\alpha$-Te structure: occupation of states just above the
Fermi level in the photoexcited material greatly reduces 
the stabilizing effect of the Peierls mechanism.

Fig. 2 shows the density of states and the bandstructure along 
the lines of highest symmetry of the hexagonal Brillouin zone, 
for \(x\) = 0.2686 and \(x\) = \(\frac{1}{3}\). 
There are nine occupied valence bands in the semiconducting structure.
As \(x\) is increased, the bandgap closes at \(x = 0.2875\), where 
the energy of the lowest conduction band at $A$ becomes equal to 
the top of the valence band at $H$. 
A direct crossing of the bands occurs for the
high symmetry, \(x = \frac{1}{3}\), structure.
\begin{figure}
\begin{center}
\includegraphics[width=10cm,height=7cm]{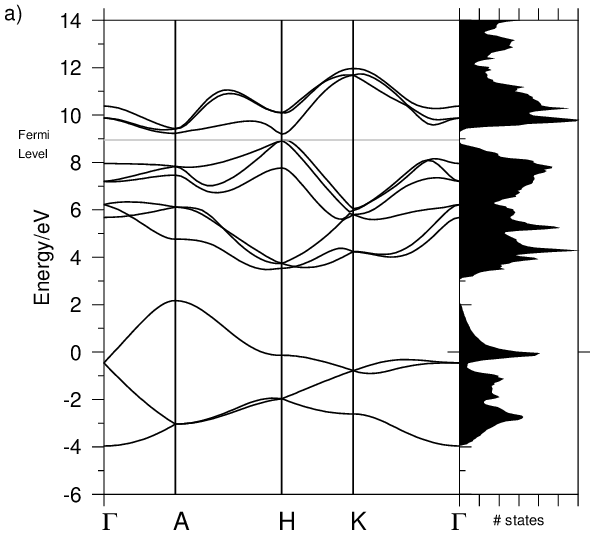}
\includegraphics[width=10cm,height=7cm]{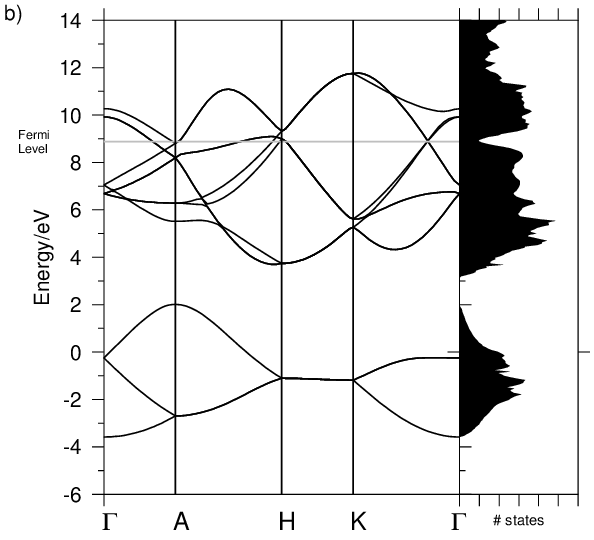}
\caption{Calculated bandstructure and density of states of %
\(\alpha\)-Te for (a) \(x=x_{equil}=0.2686\) and (b) \(x = \frac{1}{3}\).}
\end{center}
\end{figure}

Within the DECP mechanism, the initial amplitude of the phonon
motion in the photoexcited material is given by the
difference between the value of $x$ before excitation
by the pump pulse
and the equilibrium value of $x$ for the photoexcited system.
Thus, the initial amplitude of phonon motion is equal to
 \(|x_{equil}-x^{0}_{equil}|\), where 
\(x_{equil}\) is the position of the energy minimum for
the photoexcited system and 
\(x^{0}_{equil}\) is the minimum for the unexcited system. 
The energy vs. $x$ curves are significantly anharmonic and so 
the period of motion was calculated from each curve for 
amplitudes ranging between 0 and the initial DECP amplitude, 
\(|x_{equil}-x^{0}_{equil}|\).
The frequency of the vibration will vary from its initial, large 
amplitude value to its final, small amplitude value as the 
motion is damped.

\begin{figure}
\begin{center}
\includegraphics[width=10cm,height=6cm]{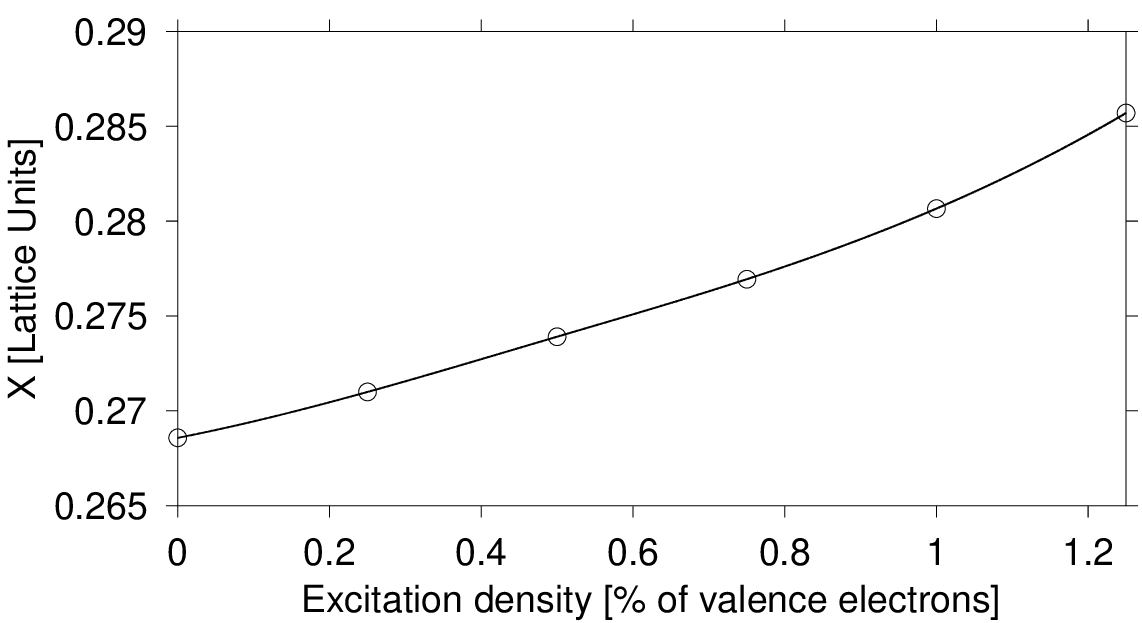}

\includegraphics[width=10cm,height=6cm]{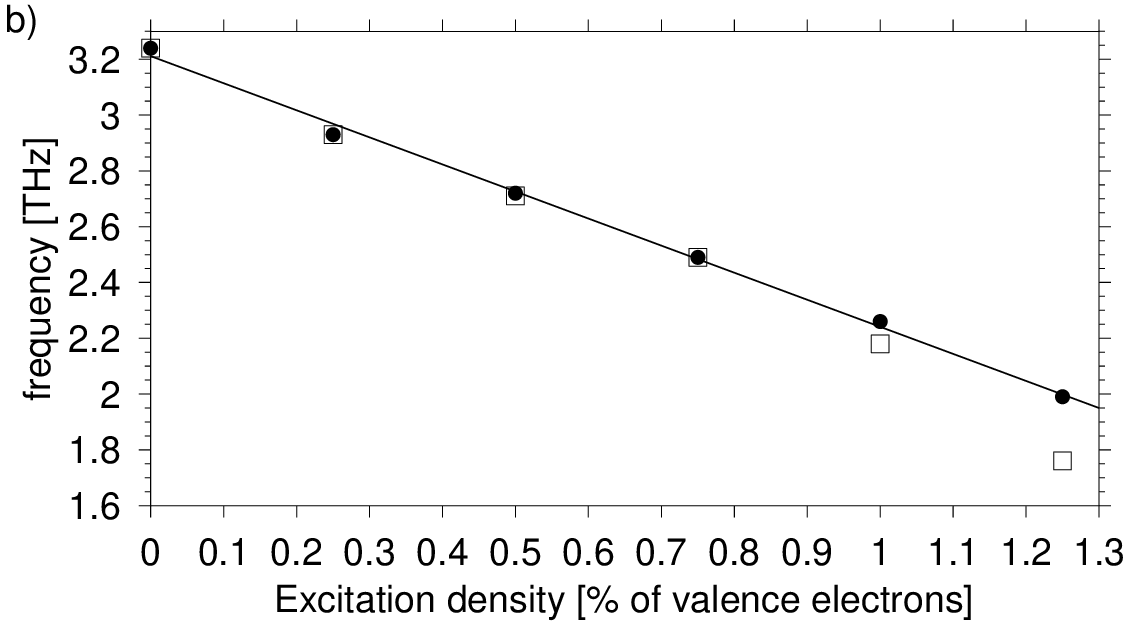}
\caption{(a) Equilibrium bondlength and (b) \(A_{1}\) phonon  %
frequency versus photoexcited carrier density.                %
The frequency is shown for the initial DECP amplitude motion  %
(open squares) and for the final, small amplitude motion (solid dots).}
\end{center}.
\end{figure}
Fig.\ 3a shows the linear increase of the equilibrium bondlength 
with carrier excitation density, 
clearly demonstrating the weakening of the bonds
and the displaced equilibrium that results from electrons 
occupying antibonding states. 
Fig.\ 3b shows the dependence of frequency on excitation
density for large and small amplitude motion. 
For all carrier densities, the anharmonic terms tend to lower
the phonon frequency.
Thus, the DECP phonon period will decrease on successive
cycles of the motion, as damping reduces the amplitude.
However, the initial amplitude of DECP motion is large
enough to make this effect appreciable only for photoexcited
carrier densities greater than approximately 1\% of the
valence electron density. 
At this level of excited carrier density, the harmonic 
phonon frequency is already reduced to 2/3 of its value in the 
system without excited carriers.

In the light of these calculations, we can offer new insight
into the recent pump-probe measurements of the $A_1$ phonon in 
tellurium.
Hunsche {\it et al.} \cite{hunsche}  measured the phonon 
frequency for pump flunces up to 12.5 mJ/cm$^2$ and found
a systematic decrease in frequency from 3.6 THz for the lowest 
pump fluences to 2.95 THz for the largest.
This reduction in phonon frequency with increasing pump fluence
is consistent with the linear decrease in phonon frequency with
photoexcited carrier density found in the calculations above.
However, using the results of Fig.\ 3b, we infer a much smaller
carrier density than one would expect, based on strictly linear 
absorption and neglecting carrier diffusion, as given in 
Ref.\ \cite{hunsche}.  For example, for a pump fluence of
2.4 mJ/cm$^2$, Hunsche {\it et al.} estimate a carrier density
of $1.1 \times 10^{21}$ cm$^{-3}$, or 0.7\% of the valence
electron density, at the sample surface.
However, for this pump fluence, the $A_1$ phonon frequency, as
measured by the probe reflectivity oscillations, is 
0.1-0.2 THz lower than its frequency in the unexcited system
(see Fig.\ 3 of Ref. \cite{hunsche}).
Using the calculated phonon frequency as a function of carrier
density (Fig.\ 3b), this frequency change would imply a carrier 
density of 0.1-0.2\% of the valence electron density.
Thus, we find that the carrier densities present in the 
experiments of Hunsche {\it et al.} are considerably smaller
than the assumption of linear absorption and neglect of carrier
diffusion would imply.
Moreover, the decrease in the period for successive oscillations
at very high pump fluence, where the phonon frequency is 0.3-0.6 THz
lower than its value in the unexcited system (Fig.\ 3 of 
Ref.\ \cite{hunsche}), is too large to be due to anharmonicity 
of the phonon motion, as shown in Fig.\ 3b for carrier densities
of approximately 0.6\% of the valence electron density. 
It is more likely that these changes in period are due to a
reduction of the carrier density caused by carrier diffusion.

In conclusion, we have calculated the dependence of the $A_1$
phonon frequency in tellurium on photoexcited carrier density,
assuming fast thermalization of the carrier distribution and slow 
electron-hole recombination.
We have also determined the initial amplitude of phonon motion, and the
resulting anharmonic corrections to the period, within
the DECP mechanism for the photoexcited system. 
The calculations shed new light on the interpretation of existing
pump-probe experiments and open the possibility of using real-time 
measurements of the $A_1$ phonon period to monitor the evolution 
of carrier densities in photoexcited tellurium on a sub-picosecond 
timescale.
Similar calculations are feasible for many other materials.
This work has been supported by Forbairt Contract SC/96/742.

\end{document}